# Stabilization of Polar Nano Regions in Pb-free ferroelectrics

A. Pramanick,[1,a)] W. Dmowski,[2,3] T. Egami,[2,3] A. Setiadi Budisuharto,[1] F. Weyland,[4] N. Novak,[4] A. D. Christianson,[5] J. M. Borreguero,[6] D. L. Abernathy,[5] MRV Jørgensen[7,8]

[1]Department of Materials Science and Engineering, City University of Hong Kong, Kowloon, Hong Kong SAR
[2]Shull Wollan Center, Oak Ridge National Laboratory, Oak Ridge, Tennessee
[3]Department of Materials Science and Engineering, University of Tennessee, Oak Ridge, Tennessee, USA
[4]Institue of Material Science, Technische Universität Darmstadt, Darmstadt, Germany
[5]Quantum Condensed Matter Division, Oak Ridge National Laboratory, Oak Ridge, Tennessee, USA
[6]Neutron Data Analysis and Visualization Division, Oak Ridge National Laboratory, Oak Ridge, Tennessee, USA
[7]Center for Materials Crystallography, iNano and Department of Chemistry, Aarhus University, Denmark
[8]MAX IV Laboratory, Lund University, Lund, Sweden

**Abstract:**
**Formation of polar nano regions through solid-solution additions are known to enhance significantly the functional properties of ferroelectric materials. Despite considerable progress in characterizing the microscopic behavior of polar nano regions, understanding their real-space atomic structure and dynamics of formation remains a considerable challenge. Here, using the method of dynamic pair distribution function, we provide direct insights into the role of solid-solution additions towards the stabilization of polar nano regions in the Pb-free ferroelectric of $Ba(Zr,Ti)O_3$. It is shown that for an optimum level of substitution of Ti by larger Zr ions, the dynamics of atomic displacements for ferroelectric polarization are slowed sufficiently, which leads to increased local correlation among dipoles below THz frequencies. The dynamic pair distribution function technique demonstrates unique capability to obtain insights into locally correlated atomic dynamics in disordered materials, including new Pb-free ferroelectrics, which is necessary to understand and control their functional properties**.

[a)]apramani@cityu.edu.hk

In materials with inhomogeneities for atomic displacements, spin directions or elastic distortions,[1-5] presence of nanoscale structural correlations, instead of a truly random structure, often leads to optimal properties. Examples include ferroelectrics,[6-9] manganites with colossal magnetoresistance,[10] superconductors,[2] spin ice[11] and strain glasses.[12] An understanding of both the local structure within nanoscale correlated regions and dynamics of their formation are necessary in order to control material properties precisely and to design new materials from a fundamental physico-chemical perspective. In spite of important theoretical and experimental advancements in this regard, gaining atomistic insights into nanoscale inhomogeneity remains a non-trivial undertaking.[13-17] Here, we demonstrate the ability to obtain structural and dynamic insights into polar nano regions (PNR) for the Pb-free ferroelectric material of $Ba(Zr,Ti)O_3$ using recent developments in the dynamic pair distribution function (DyPDF) technique.[18,19]

Ferroelectrics exhibit a spontaneous electrical polarization due to relative displacements of cations and anions. In some ferroelectrics, PNRs form when energetically degenerate displacements of atoms become correlated within nanoscale domains, but such local ordering does not extend over longer range.[8] In recent years, there has been intense scrutiny about the nature of PNRs due to their presumed role for highly enhanced dielectric and piezoelectric properties.[7,20-22] In traditional Pb-based ferroelectrics, the "static" atomic displacements within PNRs were modeled based on X-ray and neutron diffuse scattering patterns.[6,7] However, it was later shown that PNRs are intrinsically dynamic in nature [18,22,23] and appear "static" when the frequency range of fluctuating dipolar moments is within the energy resolution of the experiment. Based on inelastic neutron scattering measurement of phonons, a slowing down of PNRs was proposed for Pb-based ferroelectrics due to a localization of polar atomic vibrational modes within nanoscale regions.[20,22] Whereas such developments are noteworthy, it remains difficult to determine the particular atomic vibrations and their energetics within PNRs from phonon scattering alone, especially for the highly damped soft transverse optical (TO) modes.

In view of serious environmental concerns regarding Pb in electronic components, the search for new Pb-free ferroelectrics has intensified recently.[24-26] Indeed, promising enhancements in functional properties have been observed for solid solutions of prototypical Pb-free ferroelectrics, such as $BaTiO_3$ and $KNbO_3$,[24,27,28] which are tentatively linked to microscopic disorder in electric polarization vectors in form of PNRs.[29,30] However, direct characterization of local structure and dynamics of formation of PNRs in such materials is missing. It is noteworthy that local structure of Pb-free ferroelectrics, such as those studied here, could be substantially different from Pb-based ferroelectrics. This is due to differences in atomic bonding environments: while hybridization between B-site (Ti or Nb) cation and O plays a major role for ferroelectricity in the Pb-free ferroelectrics, Pb-O (A-site) hybridization is more significant for the Pb-based ferroelectrics.[31,32] Indeed, peculiarly for $BaTiO_3$ and $KNbO_3$, disorder in B-site atomic displacement vectors is observed even in absence of any chemical substitution,[21,33-35] unlike for Pb-based ferroelectrics. In solid-solutions of these Pb-free $ABO_3$ ferroelectrics, it was proposed that various atomic substitutions further modify the local correlations among the B-site atomic displacements leading to formation of PNRs.[36] However, the exact nature of how atomic displacements in PNRs become modified with solid-solution additions is not known at the moment. In order to rationally design new Pb-free ferroelectrics with enhanced properties, it is expected that such knowledge will be crucial. We demonstrate here direct experimental insights into the local structure and dynamics of formation of PNRs in the important Pb-free ferroelectric $Ba(Zr_xTi_{1-x})O_3$.

$Ba(Zr_xTi_{1-x})O_3$ has received great interest as an alternative to Pb-based ferroelectrics with attractive dielectric, piezoelectric and electrocaloric properties.[37,38] The phase diagram of the system is shown in Figure 1(a)[39], whereas the dielectric permittivities for different compositions are shown in 1(b). For details on synthesis and property measurements, see Supplementary Information. For $Ba(Zr_xTi_{1-x})O_3$, a transition from normal ferroelectric to relaxor behavior has been proposed within the composition range of $x$ = 0.15-0.2.[38,40] Interestingly, the maximum in dielectric permittivity is observed for $x$ = 0.15, but then drops sharply and is asymmetrical for $x$ = 0.2 and beyond. Figure 1(c) compares the ferroelectric hysteresis loops measured for ceramics of compositions $x$ = 0.10 and $x$ = 0.15. It can be observed that both the maximum ($P_{max}$) and the remnant polarization ($P_r$) are significantly larger for $x$ = 0.15 as compared to $x$ = 0.10, although they both exist in the rhombohedral phase at -50 °C. A similar trend can also be observed at 30 °C. Such behaviors imply the influence of local correlations among disordered polarization vectors on material properties. Evidence for local dipolar correlations can also be observed from various other anomalies as described below.

Figure 1(d) shows the trend for $1/\varepsilon_r$ for both compositions; the actual values are scaled for a comparison of the respective trends. For a ferroelectric with a relatively large double-well potential, Curie-Weiss behavior $1/\varepsilon_r \sim (T-T_c)$ can be expected, as indicated by the straight line, where $T_c$ is the Curie temperature. For $x$ = 0.15, significant deviation from the Curie-Weiss behavior is observed at low temperatures, more than that of $x$ = 0.10. Deviation from the Curie-Weiss behavior indicates stronger effects of local dipole-dipole correlations.[41] If dipoles are purely a result of zero-point quantum fluctuations, one observes a $1/\varepsilon_r \sim (T-T_C)^2$ behavior.[42] Since, the line for $x$ = 0.15 is closer to the $1/\varepsilon_r \sim (T-T_C)^2$ curve, we can infer that local dipolar correlations are more significant for this composition. At higher temperatures, both the compositions are close to the Curie-Weiss behavior, indicating reduced importance of local dipole-dipole correlations.

Evidence of local dipolar correlations can also be observed from temperature-dependent polarization and electrocaloric properties. As shown in Figure 1(e), for $x$ = 0.15, significant non-zero values for both $P_{max}$ and $P_r$ (see Figure S4 in Supplementary) can be observed beyond the ferroelectric-paraelectric transition temperature $T_C$ ~ 55 °C, which can be attributed to the coalescence of local polarization vectors within PNRs.[43] Recently, Ba(Zr$_x$Ti$_{1-x}$)O$_3$ ceramics were also shown to exhibit giant electrocaloric effect, or adiabatic temperature change Δ$T$ under electric-field application, which can be used for energy-efficient solid-state cooling technologies.[38] Figure 1(f) shows an intriguing behavior of electrocaloric temperature change in this material under moderate electric-field amplitudes. With increase in electric-field amplitude, along with an increase in Δ$T$, there is also a shift in the temperature for maximum Δ$T$ from ~52 °C at 0.2 kV/mm to ~62 °C at 2 kV/mm. Normally, for a ferroelectric with first-order phase transition, one can expect maximum Δ$T$ to occur at $T_C$ due to large change in ferroelectric polarization. However, a shift in the position of the maximum Δ$T$ from $T_C$ can occur, as demonstrated here, if strong entropic contribution comes from phenomena associated with PNRs.[44]

We used the Dynamic Pair Distribution Function (DyPDF) method to investigate the exact nature of local dipole-dipole correlations within PNRs. (For details see Supplementary Information). The DyPDF or $G(r,E)$ is a function of pairwise atomic distance $r$ and energy $E$, which is obtained by Fourier transformation of the normalized total scattering factor $S(Q,E)$ over scattering vector $Q$.[18,19] It reveals the local atomic distance correlations at different frequencies given by $v = E/h$. For $E$ = 0, the Fourier transform of $S(Q, E)$ gives the atomic distances of the time-averaged structure, whereas the Fourier-transform of the integrated total scattering factor over energy, $S(Q) = \int S(Q,E)dE$, gives the same-time correlations as in the conventional pair distribution function (PDF) obtained, for example, by X-ray scattering.[15] The peak positions in the PDF indicates the distances between the different atomic pairs, while the height of each peak corresponds to the probability of finding specific atomic-pair at this distance. In other words, peak intensities are higher for greater correlations among atomic positions. The DyPDF or $G(r,E)$ additionally provides information on local structure as a function of frequency or time-scales for atomic motions. This is significant since PNRs in ferroelectrics are known to have intrinsically dynamic characteristics. Additionally, real-space characterization of correlated atomic dynamics using DyPDF becomes particularly important to understand the soft phonon modes, which are usually highly overdamped and can be localized.

Figure 2(a) shows DyPDF pattern for $E$ = 0, that is, the time-averaged local structure at temperature $T$ = 63 °C. The peaks for Ti(Zr)-O and Ba-Ti(Zr) are particularly interesting, since they indicate off-centering of the B atoms. The peak for Ba-Ti/Ba-Zr bond-distance forms a shoulder next to the peak for the same-element correlations (Ba-Ba,Ti(Zr)-Ti(Zr), O-O) at 4 Å. The time-averaged structures for the two compositions are nearly identical. The PDF patterns are also close to the one reported earlier from X-ray total scattering for a similar compound.[45] It should be noted that in an X-ray experiment, $S(Q)$ is obtained by integrating over all energy transfers, however, the integrated pattern is dominated by elastic scattering and therefore the resulting PDF is similar to the time-averaged local structure. Figure 2(b) shows the DyPDFs for $E$ = 3.4 meV, for which we see a clear splitting of the peaks near 4 Å for $x$ = 0.15 but not for x = 0.10. Also, the peaks for Ti(Zr)-O have broadened and their centers moved to lower $r$. These features indicate an increased correlation among dynamic off-centering of the Ti(Zr) atoms at $E$ = 3.4 meV for $x$ = 0.15, but not for $x$ = 0.10, even though the time-averaged structures for both compositions are nearly identical. Figure 2(c) shows the DyPDFs for a higher $E$ = 6.4 meV, for which the $G(r,E)$ are again nearly identical, indicating that the dynamic off-centering of the Ti(Zr) atoms at $E$ = 6.4 meV are now similarly correlated for both the compositions. The energy range near which the peak-splitting is observed closely matches with the energy of the zone center TO modes in prototypical BaTiO$_3$, whereas the shift in atomic peak positions is ~ 0.1 Å, consistent with ferroelectric polarization,[31,32] these therefore suggests that the dynamic Ti (Zr) off-centering corresponds to the soft TO vibrational modes. While this is hard to deduce from spectroscopic measurements of phonons in $Q$-space, the DyPDF method contains direct information about the specific atomic off-centering corresponding to polar vectors in real-space. Here, we note that both Ti(Zr)-O and Ba-Ti(Zr) bond distances are affected, which indicates that the atomic off-centering occur close to the body diagonal or <111>, in agreement with several earlier studies.[29,33]

A 2-D plot of DyPDF vs $r$ or atomic-pair distance incorporating the $G(r,E)$ for all $E$s obtained for each composition at 63 °C are shown in Figures 3(a,b). The differences are clear for the two compositions. The peak splitting at ~4 Å, as described above, occurs at a lower energy level $E$ for $x$ = 0.15 than that for $x$ = 0.10. At the same time, the correlations among the dynamic atomic off-centerings at lower energies are also higher for $x$ = 0.15. This is significant, since DyPDF measures the atomic pair-wise correlations that are formed at frequencies $\nu = E/h$. $E$ = 3.4 meV and $E$ = 6.4 meV correspond to frequencies of $\nu$ = 0.8 THz and $\nu$ = 1.6 THz, respectively. The fact that the dynamic atomic off-centering occurs at lower energies for $x$ = 0.15 should therefore naturally translate to increased local correlations for such displacements in the time-domain. At sub-THz frequencies, the dynamic Ti displacements from TO vibrational modes approaches that of the hopping motion of Ti ions,[34] and therefore their resonance contributes to the stabilization of PNRs. This is consistent with the data shown in Figure 1(c), where at lower temperatures, $1/\varepsilon_r$ shows larger deviation from Curie-Weiss behavior for $x$ = 0.15. It can therefore be inferred that the higher dielectric permittivity peak as well as higher $P_r$ and $P_{max}$ for $x$ = 0.15, as shown in Figure 1(b,c), mainly originates from stabilization of PNRs through correlated atomic displacements below the THz limit. Stabilization of the PNR dynamics below THz frequencies is also the principal reason for inducing ferroelectric-to-relaxor transition near composition $x$ = 0.15, which is responsible for the large electrocaloric properties.[38] Our results here are consistent with earlier propositions that when correlated atomic vibrations within PNRs slow down below the THz limit, they can have significant effects on macroscopic properties, such as enhanced dielectric and ferroelectric properties.[46,47] This demonstrates the critical role of the solid-solution additive Zr towards enhancement of functional properties in Ba(Zr$_x$Ti$_{1-x}$)O$_3$.

Next we explore the causes for the increased correlations among the dynamic atomic off-centering for specific compositions, in this case $x$ = 0.15. At $T$ = 63 °C, $x$ = 0.15 is close to the triple-point, where four phases coexist (see Figure 1(a)). For x = 0.10, the same temperature corresponds to the boundary between the tetragonal and the orthorhombic phases. In order to examine if such phase difference affects local atomic displacements, we performed neutron scattering experiments at -50 °C (Figures 3(c,d)), where both compositions exist in the rhombohedral phase. The corresponding $G(r,E)$ patterns at 63 °C and -50 °C are strikingly similar. Therefore, we can conclude that the average phase is not the principal factor for different dynamic atomic displacements for the two compositions. However, structural disorder due to thermal activation do play an important role. At 267 °C, as shown in Figures 3(e,f), the $G(r,E)$ patterns are similar for both compositions, which indicate similar correlations among dynamic atomic displacements for the two compositions. This is consistent with the data shown in Figure 1(c), where $1/\varepsilon_r$ starts to merge at higher temperatures.

Since the average structural phase cannot explain the composition-dependence of local dynamic atomic correlations, we explore other possible factors. According to the order-disorder model, the disordered <111> Ti atomic displacements locally correlated in the form of 1-dimensional chains, which are a few nanometers long.[33] Introduction of a larger cation at the B-site, such as Zr instead of Ti, creates local stresses which can disturb the long-range transverse correlations among the 1-dimensional chains.[36] From this line of argument, we would expect that the net polarization should get lowered with progressively increasing addition of Zr. Instead, the dielectric permittivity $\varepsilon_r$ peaks at an intermediate Zr content (Figure 1(b)), indicating non-trivial compositional effects. It is known that stress centers can alter the level for double-well potential in ferroelectrics.[48] Substitution of Ti with Zr will create such centers locally, where the energy for off-center atomic displacements will be lowered and as a corollary the frequency of polar displacements at these centers will also decrease. We also note a recent work in which the authors using first-principles calculations found that addition of Zr on Ti sites leads to a flattening of the double-well potential at the phase-coexistence region near triple point.[49] Therefore, when enough of such centers are created with increased Zr addition, they can start to coalesce, leading to an overall decrease in energy for dynamic atomic displacements and consequently stabilize polar clusters in the form of PNRs below THz frequency. However, when Zr is added in excess, it causes net negative effect by creating even more disruption for the transverse correlations among 1-D chains as well as by flattening or even removal of the double-well potential. This is likely the case in Ba($Zr_x$Ti$_{1-x}$)O$_3$ for $x \geq 0.2$, where the dielectric permittivity abruptly decreases.

In summary, we used the dynamic pair distribution function (DyPDF) derived from inelastic neutron scattering to obtain direct insights into the atomic structure and dynamics of locally correlated PNRs in the Pb-free ferroelectric Ba($Zr_x$Ti$_{1-x}$)O$_3$. The findings establish that the dynamic B site atomic displacements occur along <111> and correlations among these off-center polar displacements are stabilized for optimum substitutions of Ti by larger Zr atoms. The critical role of Zr addition is to lower the energies for off-center atomic displacements, which consequently reduce the frequency of atomic hopping within the PNRs below THz frequencies. Based on this, it is proposed that future research on Pb-free ferroelectrics should focus on evaluating the role of solid-solution additions on both the structure as well as the dynamics of atomic correlations at the local level. The current methodology can be used to obtain atomistic insights into the locally correlated structures in a wide variety of disordered materials and can therefore help design materials from fundamental physico-chemical perspectives.

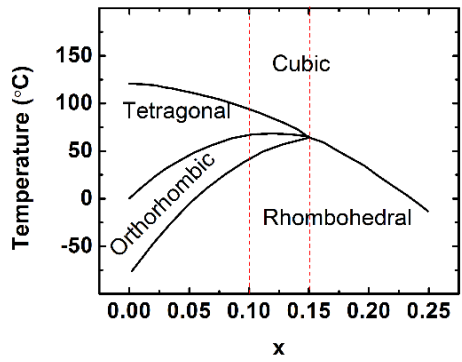

(a)

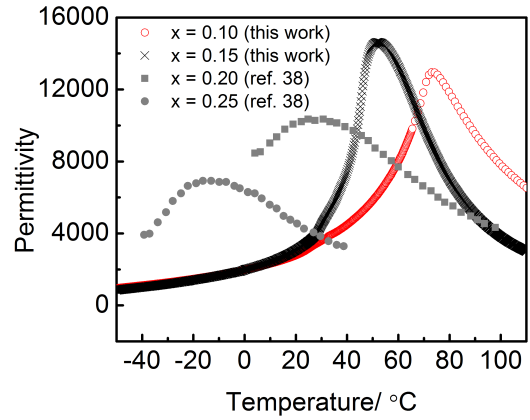

(b)

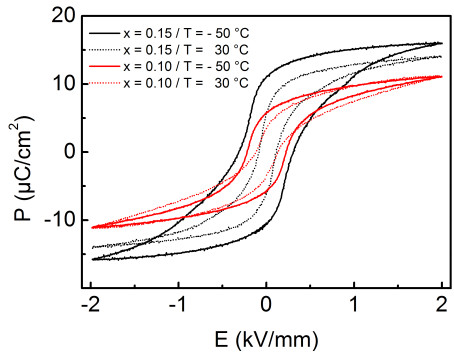

(c)

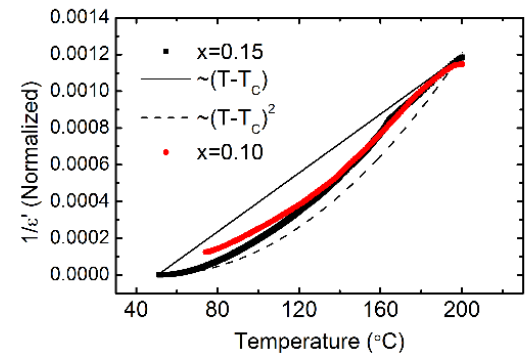

(d)

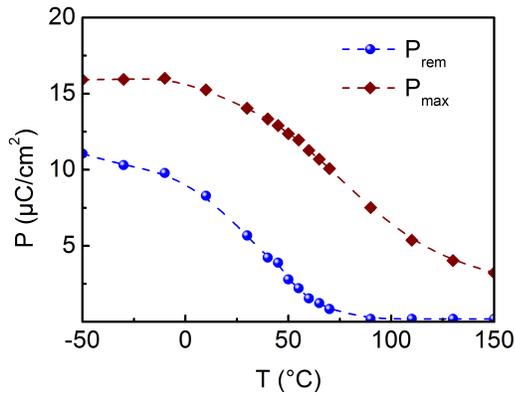

(e)

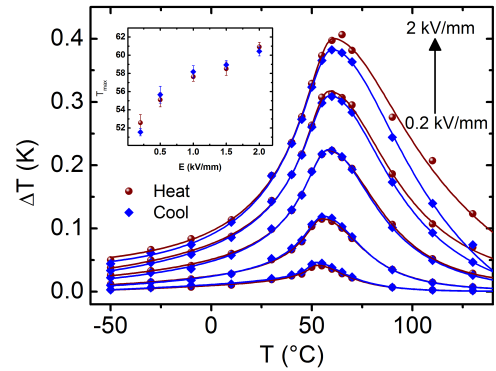

(f)

Figure 1: (a) Phase diagram of the Ba(Zr$_x$Ti$_{1-x}$)O$_3$ solid-solution.[38,39] The dotted lines indicate the specific compositions for which DyPDF characterization was undertaken. (b) The real component of dielectric permittivity $\varepsilon_r$ as function of temperature for different compositions. (c)

Comparison of *P-E* hysteresis loops for ceramics of composition *x* = 0.10 and *x* = 0.15. The decrease in coercive fields at 30 °C is likely due to thermally facilitated initiation of domain wall motion (d) $1/\varepsilon_r$ for *x* = 0.15 and *x* = 0.10, normalized for comparison. The curves are compared to expected behavior for the two exponents, while taking $T_C$ ≈ 50.5 °C as reference. (e) Temperature dependence of $P_r$ and $P_{max}$ for *x* = 0.15 ceramic display non-zero values above $T_C$ (f) Change in temperature $\Delta T$ as a result of electrocaloric effect for *x* = 0.15. The peaks are fitted with an asymmetric Pearson VII function. The temperature of maximum $\Delta T$ as a function of applied electric-field amplitude is shown in the inset.

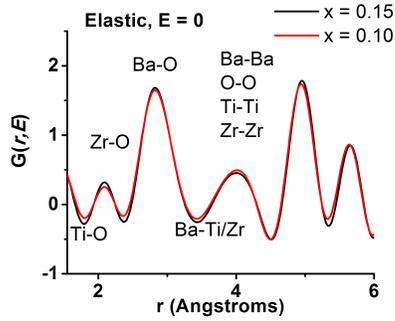

(a)

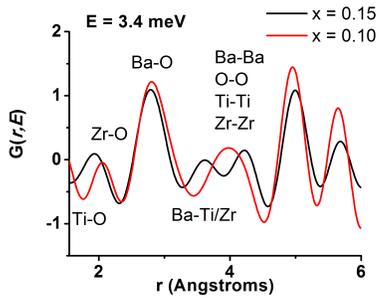

(b)

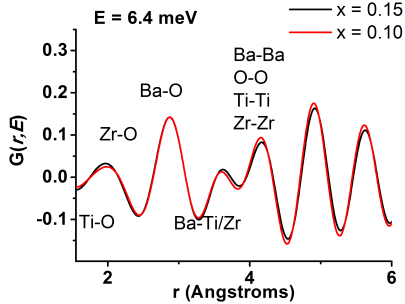

(c)

Figure 3: Comparison of DyPDF $G(r,E)$ for $x$ = 0.15 and $x$ = 0.10 at various energy transfers $E$, measured at $T$ = 63 °C. The atomic-pair bond distances ($r$) are also marked on the plot. Since the experimental data presented here are obtained using neutron scattering, the Ti-O distance will have a negative peak due to the negative scattering length of Ti. The local atomic structural correlations significantly differ for two compositions at the intermediate value of $E$ = 3.4 meV.

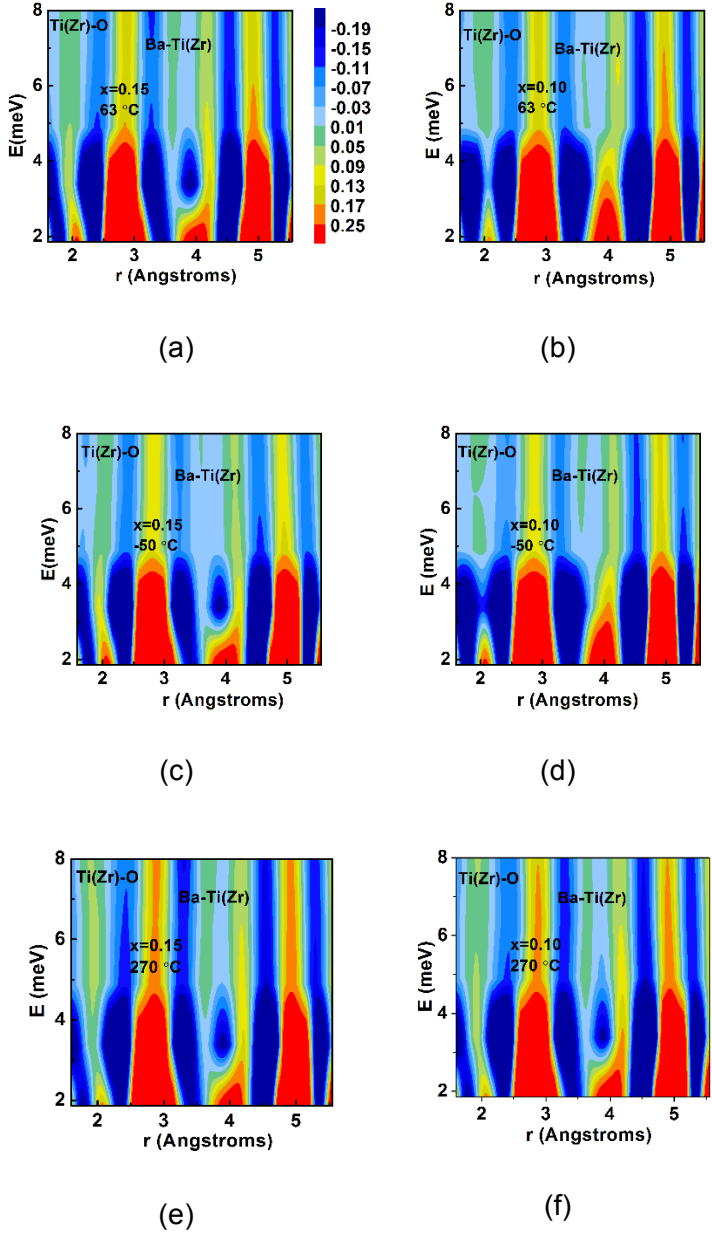

Figure 4: Comparison of composite DyPDF patterns $G(r,E)$ for Ba(Zr$_x$Ti$_{1-x}$)O$_3$ for different compositions and temperatures: (a) $x$ = 0.15, $T$ = 63 °C, (b) $x$ = 0.10, $T$ = 63 °C, (c) $x$ = 0.15, $T$ = -50 °C, (d) $x$ = 0.10, $T$ = -50 °C, (e) $x$ = 0.15, $T$ = 270 °C, (f) $x$ = 0.10, $T$ = 270 °C. The DyPDF patterns for the two compositions differ for lower temperatures of $T$ = -50 °C and 63 °C, but appears similar for higher temperature of $T$ = 270 °C.

**Acknowledgements**

AP gratefully acknowledges funding support from CityU Start-up Grant for New Faculty (Project Numbers 7200514 and 9610377). A portion of this research used resources at the Spallation Neutron Source, a DOE Office of Science User Facility operated by the Oak Ridge National Laboratory. WD, TE and JMB were supported by the U.S. Department of Energy, Office of Science, Basic Energy Sciences, Materials Science and Engineering division. M.R.V.J. is grateful for the support by the Danish National Research Foundation (DNRF93), and the Danish Research Council for Nature and Universe (Danscatt). AP gratefully acknowledges technical assistance from Mr. Daniel Yau. AP acknowledges helpful discussion on this topic with Souleymane Omar Diallo.

# Supplementary Information to Stabilization of Polar Nano Regions in Pb-free ferroelectrics

A. Pramanick,[1,a)] W. Dmowski,[2,3] T. Egami,[2,3] A. Setiadi Budisuharto,[1] F. Weyland,[4] N. Novak,[4] A. Christianson,[5] J. M. Borreguero,[6] D. L. Abernathy,[5] M. R. V. Jørgensen,[7,8]

[1]Department of Materials Science and Engineering, City University of Hong Kong, Kowloon, Hong Kong SAR
[2]Shull Wollan Center, Oak Ridge National Laboratory, Oak Ridge, Tennessee
[3]Department of Materials Science and Engineering, University of Tennessee, Oak Ridge, Tennessee, USA
[4]Institue of Material Science, Technische Universität Darmstadt, Darmstadt, Germany
[5]Quantum Condensed Matter Division, Oak Ridge National Laboratory, Oak Ridge, Tennessee, USA
[6]Neutron Data Analysis and Visualization Division, Oak Ridge National Laboratory, Oak Ridge, Tennessee, USA
[7]Center for Materials Crystallography, iNano and Department of Chemisrty, Aarhus University, Denmark
[8]MAX IV Laboratory, Lund University, Lund, Sweden

## I. Materials Synthesis

Powder samples of Ba(Zr$_x$Ti$_{1-x}$)O$_3$ were prepared using conventional solid-state reaction method. Powder samples of BaCO$_3$, ZrO$_2$ and TiO$_2$ were obtained from Alfa Aesar, which were combined in appropriate proportions following the respective stoichiometries using ball milling for 24 hours. The milled powders were calcined at 1250 °C for 5 hours. Powder X-ray diffraction patterns were obtained for the two powder samples using a laboratory X-ray diffractometer. The patterns are shown below in Figure S1. They show both the powder samples were obtained in perovskite phase.

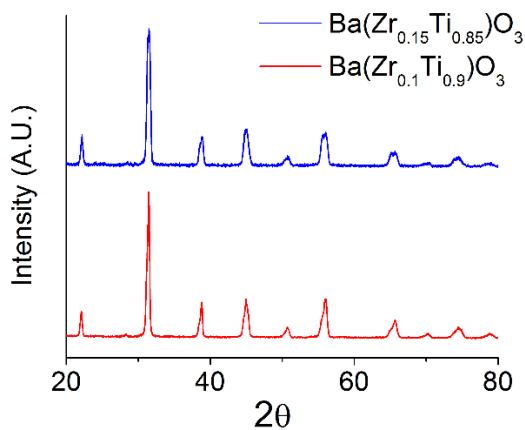

Figure S1: X-ray diffraction data from the x=0.10 and x=0.15 samples.

The calcined powders were further ball-milled to obtain powders of finer size. Following the second ball-milling, the dried powders were sintered following a two-stage process, first at 1300 °C for 4 hours and then 1500 °C for 6 hours. The density of the sintered pellets were measured to be ~94% using the Archimedes method.

## II. Dielectric and Ferroelectric Measurements

The sintered ceramics were coated with gold electrodes and contacted with electrical wires for dielectric measurements. The dielectric properties were measured in the frequency range from 1 Hz to 1MHz using an Agilent 4284A Precision LCR meter. All measurements were taken using an AC test electric field of 100 mV/mm, while heating up from -50 °C to 200 °C. No significant frequency dispersion of the dielectric permittivity maximum was observed for the samples. The results shown are measurements taken at 1000 Hz.

Ferroelectric polarization was measured by a Sawyer-Tower circuit with a reference capacitance of 16.77 µF. The electric field was linearly cycled to ±2 kV/mm (High Voltage Amplifier 20/20, Trek Inc.) with a frequency of 1 Hz by a function generator (33220A, Agilent Technologies Inc.).

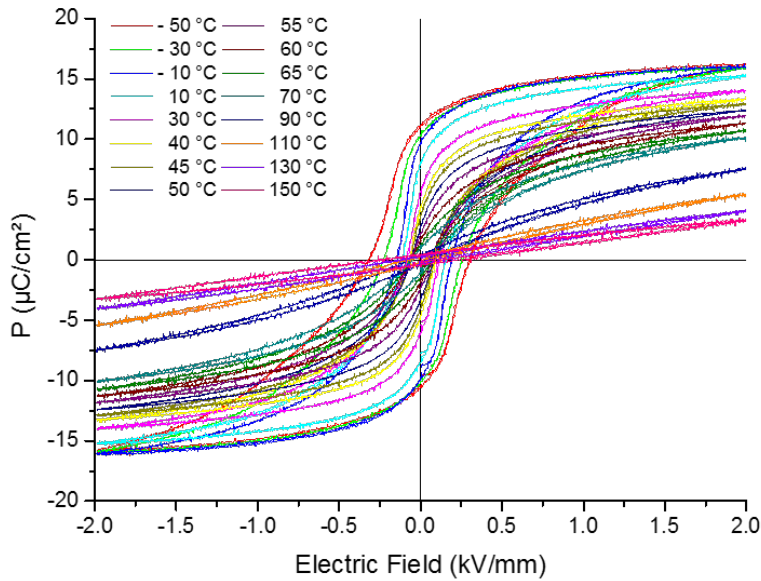

Figure S2: Polarization-electric field hysteresis loops for Ba(Zr$_{0.15}$Ti$_{0.85}$)O$_3$ as function of temperature. The $P_{max}$ and $P_r$ at each temperature (plotted in Figure 1(e)) are obtained

from the values of the maximum polarization and the remanent polarization, respectively, of the measured hysteresis loops.

## III. Electrocaloric Measurements

For the electrocaloric measurements, a small bead thermistor (GLS9, TE Connectivity Ltd.) was attached onto one surface to monitor the sample temperature. The thermistor resistance was measured by a multimeter (HP 34401A, HP Inc.). Temperature dependent electrocaloric measurements were conducted in a home-built measurement chamber. The temperature in the chamber was controlled by a temperature controller (Eurotherm 3216, Schneider Electric Systems Germany GmbH) and was stabilized within ±2 mK. Electric fields were applied/removed within 0.1 ms to ensure adiabatic conditions.

## IV. Neutron Scattering Experiments and Data Analysis

The neutron scattering measurements of the powder samples were undertaken at the ARCS spectrometer of the Spallation Neutron Source, Oak Ridge National Laboratory. Approximately 5-8 gm of powder samples were sealed in Vanadium cans, which were loaded in a cryostat. The measurements were obtained using pulsed neutrons in the time-of-flight mode. An incident energy of 70 meV for the neutrons was used, which provided the optimum range of $Q$ and $E$ for measuring relevant atomic displacement dynamics. The measured scattering patterns were normalized with respect to that from a Vanadium standard of similar dimension.

An example of the neutron scattering data from one of the powder samples used in this study is shown in Figure S3. The data was collected with an incident beam energy of 70 meV and normalized with respect to scattering signal from a Vanadium rod collected with white incident beam. The data is shown as function of scattering vector $Q$ and energy transfer $E$.

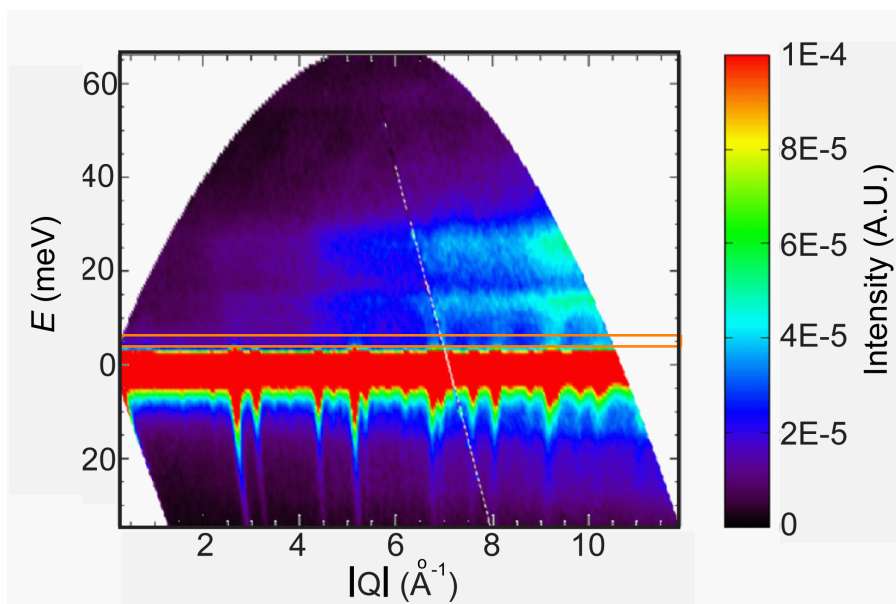

Figure S3: Neutron scattering data measured for Ba(Zr$_{0.15}$Ti$_{0.85}$)O$_3$ at ARCS with an incident beam energy of 70 meV.

The data was subsequently segmented into slices of ~ 1.5 meV width, as indicated by the orange colored box. The scattering pattern obtained after integration over a specific energy-transfer range, as shown in the top panel of Figure S4, was subsequently treated with the DyPDF Background Removal Program in Mantid.[S1] Here a quadratic background function was used, see Figure S4. The difference between the measured scattering intensity and the background function is then plotted as $S(Q,E)$, as shown in the lower panel of Figure S4.

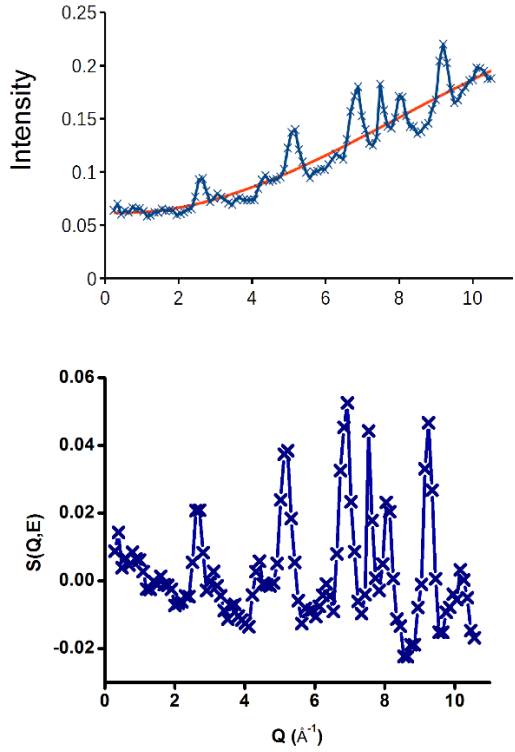

Figure S4: The top panel shows the integrated scattering pattern for median energy transfer of 4.89 meV. The background is fitted with a quadratic function to obtain the corresponding $S(Q,E)$ pattern, as shown in the lower panel.

The $S(Q,E)$, obtained after background subtraction, were then used for Fourier transformation to obtain $G(r,E)$ following

$$G(r,E) = \left(\frac{2}{\pi}\right) \int_0^{Q_{max}} S(Q,E)\sin(Qr)dQ$$

A damping function was used to ensure that $S(Q_{max}) \to 0$ in the Fourier transformation routine. The background-subtracted patterns $S(Q,E)$ were used for Fourier transformation to obtain $G(r,E)$ using a code written in IGORPro.

For all neutron scattering measurements performed at ARCS, the powders were sealed in vanadium cans. The DyPDF or $G(r,E)$ patterns generated from the empty vanadium cans using the procedure outlined above did not show any features beyond the range for elastic scattering (<3 meV), see Figure S5. This is expected for vanadium since the incoherent scattering pattern dominates over coherent scattering from phonons or lattice vibration modes. This result essentially confirms that no artifacts from instrument or data treatment procedure is introduced in the reported measurements.

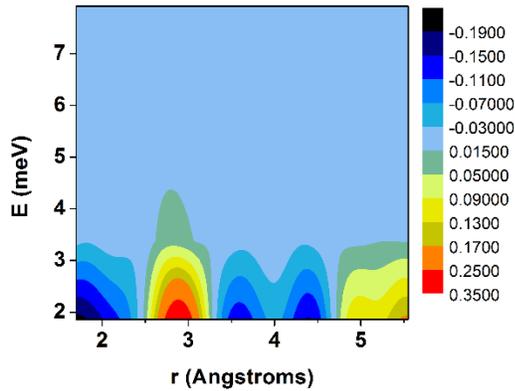

Figure S5: The DyPDF $G(r,E)$ pattern for empty vanadium can measured at 63 °C.

Characterization of time-resolved local atomic structure from energy-resolved dynamic scattering pattern $S(Q,E)$ for the studied sample Ba(Zr$_x$Ti$_{1-x}$)O$_3$ is illustrated for the composition x =0.15 in Figure S6. Fourier transformation of the elastic pattern provides the time-averaged local structural correlations $G(r,E)$, where $E=0$, while Fourier transformation of $S(Q,E)$ for energy-transfer of $E$ provides local structural correlations for frequency $v = E/h$, or for timescale $\sim h/E$. Illustrative data for composition x = 0.15.

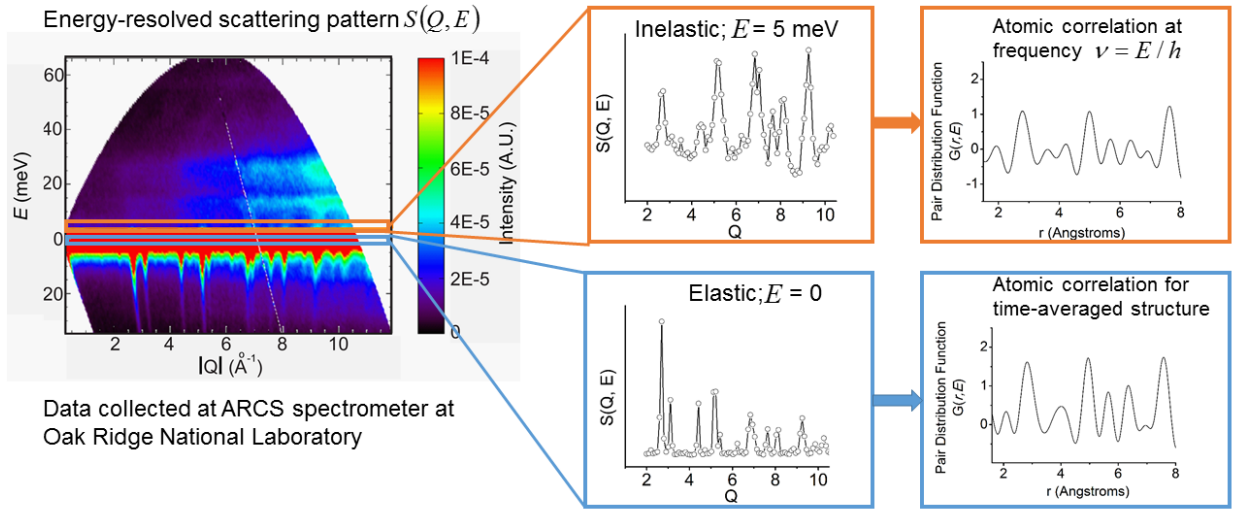

Figure S6: Characterization of time-resolved local atomic structure from energy-resolved dynamic scattering pattern $S(Q,E)$.